\documentclass[twocolumn,showpacs,preprintnumbers,amsmath,amssymb,prl,superscriptaddress]{revtex4-1}
\usepackage{graphicx}
\usepackage{dcolumn}
\usepackage{mathrsfs}
\usepackage{bm}
\usepackage{color}
\usepackage{lipsum}

\begin{document}

\preprint{}

\title{Einstein-Podolsky-Rosen Energy-Time Entanglement of Narrowband Biphotons}

\author{Yefeng Mei} 
\affiliation{Guangdong Provincial Key Laboratory of Quantum Engineering and Quantum Materials, GPETR Center for Quantum Precision Measurement and SPTE, South China Normal University, Guangzhou 510006, China}
\affiliation{Department of Physics, The Hong Kong University of Science and Technology, Clear Water Bay, Kowloon, Hong Kong, China}
\affiliation{William Mong Institute of Nano Science and Technology, The Hong Kong University of Science and Technology, Clear Water Bay, Kowloon, Hong Kong, China}

\author{Yiru Zhou} 
\affiliation{Guangdong Provincial Key Laboratory of Quantum Engineering and Quantum Materials, GPETR Center for Quantum Precision Measurement and SPTE, South China Normal University, Guangzhou 510006, China}

\author{Shanchao Zhang}\email{sczhang@m.scnu.edu.cn} 
\affiliation{Guangdong Provincial Key Laboratory of Quantum Engineering and Quantum Materials, GPETR Center for Quantum Precision Measurement and SPTE, South China Normal University, Guangzhou 510006, China}

\author{Jianfeng Li}
\affiliation{Guangdong Provincial Key Laboratory of Quantum Engineering and Quantum Materials, GPETR Center for Quantum Precision Measurement and SPTE, South China Normal University, Guangzhou 510006, China}

\author{Kaiyu Liao}
\affiliation{Guangdong Provincial Key Laboratory of Quantum Engineering and Quantum Materials, GPETR Center for Quantum Precision Measurement and SPTE, South China Normal University, Guangzhou 510006, China}

\author{Hui Yan} \email{yanhui@scnu.edu.cn}
\affiliation{Guangdong Provincial Key Laboratory of Quantum Engineering and Quantum Materials, GPETR Center for Quantum Precision Measurement and SPTE, South China Normal University, Guangzhou 510006, China}

\author{Shi-Liang Zhu} \email{slzhu@nju.edu.cn}
\affiliation{National Laboratory of Solid State Microstructures, School of Physics, Nanjing University, Nanjing 210093, China}
\affiliation{Guangdong Provincial Key Laboratory of Quantum Engineering and Quantum Materials, GPETR Center for Quantum Precision Measurement and SPTE, South China Normal University, Guangzhou 510006, China}

\author{Shengwang Du}\email{dusw@ust.hk}
\affiliation{Department of Physics, The Hong Kong University of Science and Technology, Clear Water Bay, Kowloon, Hong Kong, China}
\affiliation{William Mong Institute of Nano Science and Technology, The Hong Kong University of Science and Technology, Clear Water Bay, Kowloon, Hong Kong, China}
\affiliation{Guangdong Provincial Key Laboratory of Quantum Engineering and Quantum Materials, GPETR Center for Quantum Precision Measurement and SPTE, South China Normal University, Guangzhou 510006, China}

\date{\today}

\begin{abstract}
{We report the direct characterization of energy-time entanglement of narrowband biphotons produced from spontaneous four-wave mixing in cold atoms. The Stokes and anti-Stokes two-photon temporal correlation is measured by single-photon counters with nano second temporal resolution, and their joint spectrum is determined by using a narrow linewidth optical cavity. The energy-time entanglement is verified by the joint frequency-time uncertainty product of $0.063\pm 0.0044$, which does not only violate the separability criterion but also satisfies the continuous variable Einstein-Podolsky-Rosen steering inequality.}
\end{abstract}


\maketitle


\begin{figure*}
\includegraphics[width=18cm]{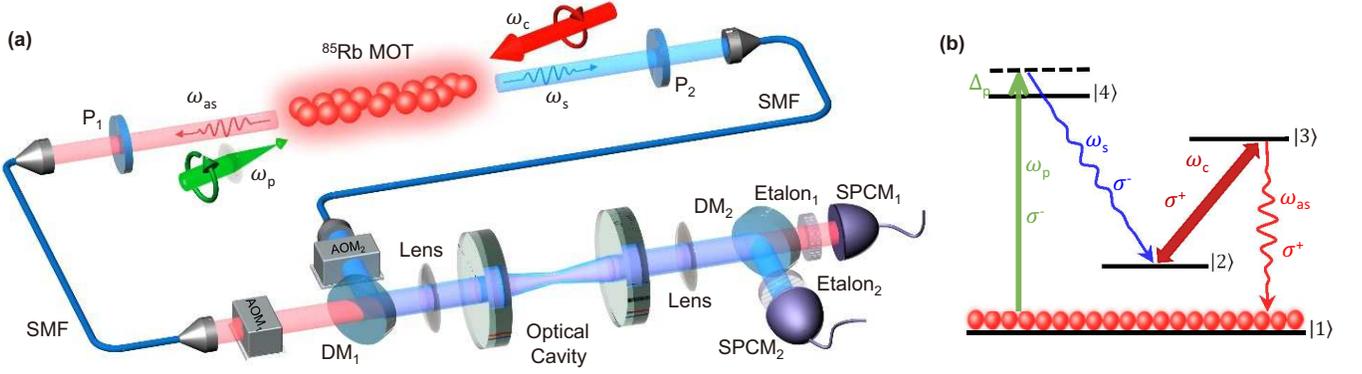}
\centering
\caption{\label{fig:schematic} Narrowband energy-time entangled biphoton generation and characterization. (a) Experiment setup. In presence of the pump ($\omega_p$) and coupling ($\omega_c$) laser beams, counter-propagating paired Stokes ($\omega_{s}$) and anti-Stokes ($\omega_{as}$) photons are produced via spontaneous four-wave-mixing (SFWM) process from a cold $^{85}$Rb atomic ensemble in a 2D magneto-optical trap (MOT). The photon pairs are then coupled into a pair of opposing single-mode fibres (SMF). To determine the joint spectral distribution, the Stokes and anti-Stokes photon frequencies are shifted by two Acousto-optical modulators (AOM$_{1}$ and AOM$_{2}$) before they are injected into the high-finesse optical cavity. Two diachronic mirrors (DM$_1$ and DM$_2$) are employed to combine and separate these photons. The transmitted photon pairs are detected by two single-photon counting modules (SPCMs). Two etalon filters are inserted in front of the SPCMs to further filter out the scattering from the pump and coupling laser beams. (b) $^{85}$Rb atomic energy level diagram for the SFWM process. $|1\rangle=|5S_{1/2}, F=2\rangle$, $|2\rangle=|5S_{1/2}, F=3\rangle$, $|3\rangle=|5P_{1/2}, F=3\rangle$, and $|4\rangle=|5P_{3/2}, F=3\rangle$.}
\end{figure*}

In the original Einstein-Podolsky-Rosen (EPR) paradox, position and momentum are taken as complementary or canonically conjugate variables to question the physical reality of quantum mechanics \cite{Einstein1935}. Now it has been clear that quantum entanglement plays an important role in the EPR nonlocality and is incompatible with the local hidden-variable theories \cite{Ou1992, Howell2004, Lee2016, Chen2019}. For more than 80 years, the EPR thought experiment has motivated many important discoveries of quantum entanglement, such as Bell's theorem \cite{Bell1964, Clauser1969}, quantum teleportation \cite{BoschiPRL1998, PanNP2010}, and quantum key distribution \cite{BB84}, as well as their quantum information applications \cite{QI}.

Time and energy are also known complementary continuous variables, whose relation is analog to that of position and momentum. The EPR energy-time entangled photon pairs (termed \textit{biphoton}s) are of great interest for long-distance quantum communication because of their insensitivity to the birefringence effect of fibers \cite{ZhangOE2008}. For a single photon state, the uncertainties of its optical angular frequency $\omega$ and arriving time $t$ are bounded by the Heisenberg's uncertainty relation $\Delta\omega \Delta t \geq 1/2$, where $\Delta$ represents standard deviation. As a result, if two photons are in a separable state, their joint frequency-time uncertainty product follows \cite{DuanPRL2000, TombesiPRL2002}
\begin{equation}
\Delta (\omega_1+\omega_2) \Delta(t_2-t_1)\geq 1.
\label{eq:Separability}
\end{equation}
Therefore, violation of the above separability criterion implies entanglement. A sufficient and more restrict condition for rising EPR paradox to rule out the local realism is the steering inequality \cite{Jones2007, Wiseman2007, Cavalcanti2009}
\begin{equation}
\Delta (\omega_1+\omega_2) \Delta(t_2-t_1) < \frac{1}{2}.
\label{eq:Steering}
\end{equation}
In fact, for a two-photon energy-time entangled state, there is no lower limit for the joint frequency-time uncertainty product. For wideband biphotons generated from spontaneous parametric down conversion (SPDC) \cite{SPDC}, while their joint spectrum is easily resolved, it is a challenge to directly and precisely measure their ultra-short temporal correlation using commercially available single-photon counters with ns resolution. Recently, direct characterization of ultrafast energy-time entangled photon pairs was demonstrated with single-photon spectrometers and sum-frequency generation correlator, giving the joint frequency-time uncertainty product of $0.290\pm0.007$ \cite{ReschPRL2018}. To our best knowledge, this is the smallest value on record for energy-time entanglement, which violates the separability criterion and satisfies the EPR steering inequality.

On the other side, narrowband biphotons with long coherence time have important applications in quantum information processing, such as realizing efficient photon-atom quantum interfaces \cite{Wang2019}. However, their EPR joint frequency-time uncertainty product has not been directly characterized because of the difficulty in resolving their ultra narrow nonclassical frequency correlation. In this Letter, we report the direct characterization of energy-time entanglement of narrowband biphotons produced from spontaneous four-wave mixing (SFWM). Owning to the long coherence time of these biphotons, the temporal correlation and uncertainty are resolved with commercial single-photon counting modules (SPCM). We make use of a narrow linewidth (72 kHz) optical cavity frequency filter to precisely map the joint spectrum and measure their energy (frequency) uncertainties. Our result of joint frequency-time uncertainty product of $0.063\pm 0.0044$ is significantly smaller than the previously reported values \cite{ReschPRL2018} and pushes its lower bound a step closer to zero.

We produce energy-time entangled narrowband biphotons from SFWM \cite{DuJOSAB2008, DuPRL2008, Zhao2014} in a cloud of $^{85}$Rb atoms prepared in a dark-line two-dimensional (2D) magneto-optical trap (MOT) \cite{Zhang2012}, as depicted in Fig.~\ref{fig:schematic}(a).  The longitudinal length of atomic cloud is around 2.0 cm and the atomic temperature is around 20 $\mu$K. The whole experiment is run periodically with a cycle duration of 1.25 ms including a MOT loading time of 0.75 ms and biphoton generation time of 0.5 ms. After the MOT loading time, the atoms are optically pumped to the ground state $|1\rangle$, as illustrated in the energy level diagram in Fig.~\ref{fig:schematic}(b). The SFWM process is driven by a pair of counter-propagating pump ($\omega_p$) and coupling ($\omega_c$) laser beams, aligned with an angle of $\theta=2.36^o$ to the longitudinal axis of the 2D MOT.  The circularly ($\sigma^-$) polarized pump laser beam (2.1 mW, 780 nm) is blue detuned by $\Delta_p=2\pi\times473~\rm MHz$ from transition $|1\rangle\rightarrow|4\rangle$ and is focus to the center of the MOT with a $1/e^2$-intensity waist diameter of $870~\rm \mu m$. The coupling laser beam ($\sigma^+$, 2.1 mW, 795 nm) is on resonance to transition $|2\rangle\rightarrow|3\rangle$, with a collimated beam diameter of 1.5 mm that covers the whole atomic cloud. Backward energy-time entangled Stokes ($\omega_s$, 780 nm) and anti-Stokes ($\omega_{as}$, 795nm) photon pairs are generated spontaneously, and collected by a pair of opposing single mode fibers (SMFs) placed along the MOT longitudinal axis. The atomic optical depth in the anti-Stokes transition is 82.

With the continuous-wave (cw) pump and coupling laser fields, the spontaneously generated phase-matched Stokes and anti-Stokes photon pairs are frequency/energy-time entangled because of the energy conservation $\omega_{as}+\omega_{s}=\omega_c+\omega_p$ originating from time translation symmetry. Therefore in an ideal SFWM case, $\Delta(\omega_{as}+\omega_{s})=0$ and the joint uncertainty product $\beta=\Delta(\omega_{as}+\omega_{s})\Delta(t_{as}-t_{s})$ approaches zero. In reality,  $\Delta(\omega_{as}+\omega_{s})$ is bounded to the finite linewidths of the pump and coupling lasers, which lead to non-zero joint uncertainty product.

We first determine the joint temporal uncertainty by measuring the two-photon temporal correlation using two SPCMs (Excelitas, SPCM-AQRH-16-FC) directly after the SMFs. In the Heisenberg picture with cw pump and coupling laser fields, ignoring linear loss and gain in the anti-Stokes and Stokes channels, the field operators at the two output surfaces can be expressed as  \cite{Zhao2016}
\begin{eqnarray}
\hat{a}_{as}(\delta \omega_{as})&=&A(\delta \omega_{as})\hat{a}_{as,0}+B(\delta \omega_{as})\hat{a}^{\dagger}_{s,0}, \nonumber \\
\hat{a}^{\dagger}_{s}(\delta \omega_{s})&=&C(\delta \omega_{s})\hat{a}_{as,0}+D(\delta \omega_{s})\hat{a}^{\dagger}_{s,0},
\label{eq:OutputFields}
\end{eqnarray}
where $\hat{a}_{as,0}$ and $\hat{a}^{\dagger}_{s,0}$ are the input vacuum fields. $\delta \omega_{as}=\omega_{as}-\bar{\omega}_{as}$ and $\delta \omega_{s}=\omega_{s}-\bar{\omega}_{s}$ are the anti-Stokes and Stokes frequency variables, with $\bar{\omega}_{as}$ and $\bar{\omega}_{s}$ as their central frequencies that satisfy $\bar{\omega}_{as}+\bar{\omega}_{s}=\omega_c+\omega_p$. Without contribution from Langevin noises, the parameters $A$, $B$, $C$, and $D$ satisfy $|A|^2-|B|^2=|D|^2-|C|^2=1$. The two-photon coincidence rate can be modeled as the Glauber correlation function in time domain
\begin{eqnarray}
T_{s,as}^{(2)}(t_{as}-t_{s})&=&\langle\hat{a}^{\dagger}_{as}(t_{as})\hat{a}^{\dagger}_{s}(t_s)\hat{a}_{s}(t_s)\hat{a}_{as}(t_{as})\rangle \nonumber \\
&=&|\psi(t_{as}-t_s)|^2+R_s R_{as},
\label{eq:Gtemporal}
\end{eqnarray}
where
\begin{eqnarray}
\psi(\tau)=\frac{1}{2\pi}\int B(\delta\omega) D^{*}(-\delta\omega) e^{-i\delta\omega\tau} d \delta\omega
\label{eq:BiphotonWavefunction}
\end{eqnarray}
is the relative biphoton temporal wavefunction. $R_{as}=\frac{1}{2\pi}\int |B(\delta\omega)|^2 d \delta\omega$ and $R_{s}=\frac{1}{2\pi}\int |C(\delta\omega)|^2 d \delta\omega$ are the anti-Stokes and Stokes single-photon rates. The first term in Eq. (\ref{eq:Gtemporal}) is the biphoton correlation, while the second term represents their accidental coincidence between different photon pair events generated spotaneously. The joint relative-time ($\tau=t_{as}-t_{s}$) uncertainty is determined by
\begin{eqnarray}
\Delta(t_{as}-t_{s})=\sqrt{\langle(\tau-\bar{\tau})^2\rangle}=\frac{\int(\tau-\bar{\tau})^2|\psi(\tau)|^2 d\tau}{\int|\psi(\tau)|^2 d\tau},
\label{eq:Deltat-t}
\end{eqnarray}
where $\bar{\tau}=\int\tau|\psi(\tau)|^2 d\tau/\int|\psi(\tau)|^2 d\tau$.

\begin{figure}
\includegraphics[width=8.6cm]{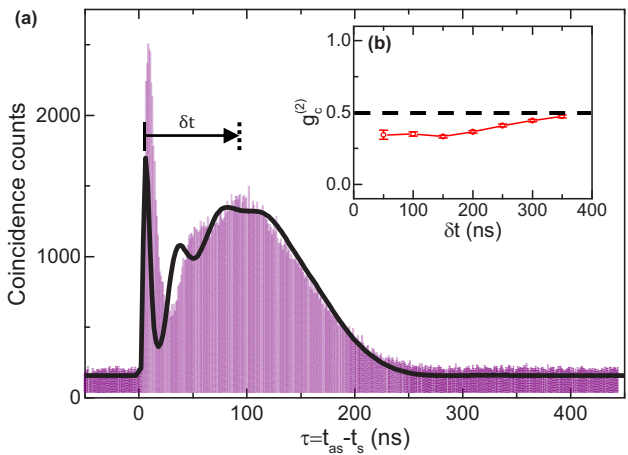}
\centering
\caption{\label{fig:temporal} Temporal correlation of energy-time entangled biphotons. (a) Two-photon temporal correlation measured as coincidence counts with a time bin of 1 ns and total data collection time 60 s. (b) The conditional autocorrelation $g^{(2)}_c$ of the heralded anti-Stokes photons as a function of the integration time width $\delta t$.}
\end{figure}

Figure \ref{fig:temporal}(a)  shows the measured Stokes and anti-Stokes photon coincidence counts, $\eta T_{s,as}^{(2)}(\tau)\Delta T t_{bin}$, where $\eta=0.034$ is the two-photon joint detection efficiency accounting the duty cycle, fiber coupling and filter transmission efficiencies, $\Delta T=60$ s is the acquisition time, and $t_{bin}=1$ ns is the time bin width. The measured coincidence counts imply that we produce biphotons from cold atoms 3,088 pairs per second. The black theoretical curve \cite{supp}, plotted from Eqs. (\ref{eq:Gtemporal}) and (\ref{eq:BiphotonWavefunction}) agrees well with the experimental data. Following Eq.~(\ref{eq:Deltat-t}), we obtain $\Delta(t_{as}-t_{s})=62.77\pm1.68$ ns. The nonclassical property of the biphoton temporal correlation is also verified by examining whether the Cauchy-Schwartz inequality \cite{ClauserPRD1974} is violated. For the data presented in Fig.~\ref{fig:temporal}(a), the normalized cross correlation function $g_{s,as}^{(2)}(\tau)=T_{s,as}^{(2)}(\tau)/(R_{s}R_{as})$ has a peak value of 15.8. With the measured autocorrelations $g_{s,s}^{(2)}(0)=g_{as,as}^{(2)}(0)=2$, the Cauchy-Schwartz inequality  $[g^{(2)}_{s,as}(\tau)]^2/[g^{(2)}_{s,s}(0)g^{(2)}_{as,as}(0)]\leq1$ is violated by a factor of 62.4. To further characterize the quantum nature of the photon pairs, we measure the conditional autocorrelation function $g^{(2)}_{c}$  \cite{Grangier1986} of anti-Stokes photons heralded by the Stokes photons. The measured $g^{(2)}_{c}$ as a function of the integration time window $\delta t$ is displayed in Fig.~\ref{fig:temporal}(b), where $g^{(2)}_{c}$ over $\delta t$ up to 300 ns is well below the two-photon threshold value of 0.5.

We now turn to the two-photon joint spectrum. Considering that the pump and coupling lasers have independent Gaussian line shapes with standard spectral deviations $\sigma_p$ and $\sigma_c$, respectively, the two-photon spectral correlation is
\begin{eqnarray}
&S_{s,as}^{(2)}(\delta\omega_s, \delta\omega_{as})=\langle\hat{a}^{\dagger}_{as}(\delta\omega_{as})\hat{a}^{\dagger}_{s}(\delta\omega_s)\hat{a}_{s}(\delta\omega_s)\hat{a}_{as}(\delta\omega_{as})\rangle \nonumber \\
&=|B(\delta\omega_{as})D(\delta\omega_s)|^2 \frac{1}{\sqrt{2\pi\sigma^2_{pc}}}\exp\big[\frac{-(\delta\omega_{as}+\delta\omega_{s})^2}{2\sigma^2_{pc}}\big] \nonumber \\
&+|B(\delta\omega_{as})C(\delta\omega_s)|^2
\label{eq:Gspectral}
\end{eqnarray}
where $\sigma^2_{pc}=\sigma_p^2+\sigma_c^2$ stands for two-photon energy fluctuation. The first term in Eq. (\ref{eq:Gspectral}) results from the frequency anti-correlation, and the second term is the accidental correlation.  We take a high-finesse ($F$=20000) optical cavity as an optical frequency filter to measure the joint spectral intensity. The optical cavity filter has a bandwidth of 72 kHz and resonance power transmission of 30\%. As shown in Fig.~\ref{fig:schematic}(a), before entering the optical cavity, Stokes and anti-Stokes photons are frequency shifted by two independently controlled acousto-optical modulators (AOMs) and then combined at a diachronic mirror (DM$_1$). With a spatial mode matching lens, the Stokes and anti-Stokes photons are coupled to the optical cavity with fixed cavity resonance frequency. The transmitted photons are separated by another diachronic mirror (DM$_2$) and then detected by the two SPCMs. Two wide-band (250 MHz) etalon filters are inserted to suppress the photon cross talk and filter out the scattering from the pump and coupling laser beams.

\begin{figure}
\includegraphics[width=8.6cm]{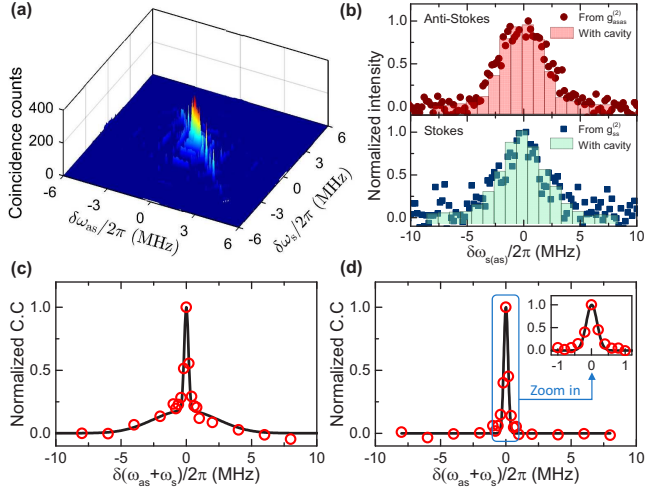}
\centering
\caption{\label{fig:spectral} Joint spectral distribution of Stokes and anti-Stokes photons. (a) The 2D joint spectral distribution measured as spectral coincidence counts between Stokes and anti-Stokes photons transmitted through the optical cavity. Each data point is taken with a two-photon coincidence time window 3 $\mu$s and acquisition time 3000 s. (b) The single-photon power spectrums of anti-Stokes and Stokes photons. The histogram data are directly measured using the high-finesse optical cavity. The dot data are calculated from the autocorrelation measurements. (c) The normalized spectral distribution as a function of the two-photon frequency sum $\delta(\omega_{as}+\omega_s)$. (d) The normalized two-photon joint spectrum with the accidental coincidence counts subtracted. The inset is the zoom-in for the frequency sum range of $\pm1$ MHz.}
\end{figure}

Shifting the frequencies of Stokes and anti-Stokes photons with the two AOMs, we obtain the joint spectral coincidence distribution, $S_{s,as}^{(2)}(\delta\omega_s, \delta\omega_{as})$, as  displayed in Figure~\ref{fig:spectral}(a). The single photon power spectrum $|B(\delta\omega_{as})|^2$ and $|C(\delta\omega_s)|^2$ can be obtained by projecting the joint spectral distribution along either $\delta\omega_{s}$ or $\delta\omega_{as}$ axis, as presented in histogram bars in Fig.~\ref{fig:spectral}(b). The Stokes (anti-Stokes) photon power spectrum can also be obtained from inverse Fourier transform of its autocorrelation functions $g^{(2)}_{s,s}(\tau)$ ($g^{(2)}_{as,as}(\tau)$), as shown as the solid dots in Fig.~\ref{fig:spectral}(b), agreeing well with the direct measurement with cavity. Both Stokes and anti-Stokes photons have similar spectral uncertainties $\Delta\omega_{as}\simeq\Delta\omega_s\simeq 2\pi\times$1.826 MHz. The 2D spectral intensity shows that the Stokes and anti-Stokes photon frequencies are anti-correlated and indicates the energy entanglement. Projecting the 2D spectral intensity along -45$^o$ direction, we get the two-photon sum frequency $\delta(\omega_{as}+\omega_s)$ distribution  and the result is shown in Fig.~\ref{fig:spectral}(c). As predicted in Eq. (\ref{eq:Gspectral}), there are two distinct features in the two-photon sum spectrum: a wide background floor from accidental coincidence counts between uncorrelated frequencies [the second term in Eq. (\ref{eq:Gspectral})], and a central narrow peak from the frequency entanglement [the first term in Eq. (\ref{eq:Gspectral})]. The solid theoretical curve agrees well with the experimental data. With the accidental coincidence counts subtracted as illustrated in Fig.~\ref{fig:spectral}(d), we fit the profile with Gaussian shape and obtain the two-photon frequency sum uncertainty $\Delta(\omega_{as}+\omega_{s})=2\pi\times(161.78\pm6.87)$ kHz. Therefore we estimate the Schmidt number of our biphoton source to be $K=1/\sqrt{1-1/\left\{1+\left[\Delta(\omega_{as}+\omega_{s})/\Delta\omega_{s/as}\right]^{2}\right\}^{2}}=8.03$, which confirms the entanglement in frequency domain while a speparable qunatum state takes $K=1$\cite{Zielnicki2018}.

Combining the result from the previous temporal measurement, we get the joint frequency-time uncertainty product
\begin{eqnarray}
\Delta(\omega_{s}+\omega_{as})\Delta(t_{as}-t_{s})=0.063\pm0.0044,
\label{eq:JointProduct}
\end{eqnarray}
which obviously violates the separability criterion in Eq.(\ref{eq:Separability}) by 212 standard deviations and thus confirms the energy-time entanglement. Moreover, the joint frequency-time uncertainty product also satisfies the steering inequality in Eq.(\ref{eq:Steering}) , which is sufficient for raising EPR paradox \cite{Jones2007, Wiseman2007, Cavalcanti2009}.

In summary, we directly characterize the energy-time entanglement of the narrowband biphotons generated from SFWM in cold atoms. The two-photon temporal correlation is measured by coincidence counts with SPCMs, and their joint spectrum is determined by a narrowband transmission optical cavity. As the Stokes and anti-Stokes photons are frequency anti-correlated, their two-photon sum frequency uncertainty $\delta(\omega_{as}+\omega_s)/(2\pi)=161.78$ kHz is much narrower than the spectral uncertainties 1.826 MHz of individual photons. We obtain the joint frequency-time uncertainty product of $0.063\pm0.0044$, which confirms the EPR entanglement and paradox. As compared to wideband photon sources, these narrowband energy-time entangled biphotons are more suitable for long-distance quantum steering \cite{Wiseman2007}, and thus will have important applications in quantum communication and quantum network based on quantum nonlocality.

\begin{acknowledgments}
The work was supported by the Hong Kong Research Grants Council (Project No. 16304817), the William Mong Institute of Nano Science and Technology (Project No. WMINST19SC05), the National Key Research and Development Program of China (Grants No. 2016YFA0302800 and No. 2016YFA0301803), the National Natural Science Foundation of  China (Grants No. 61378012, No. 91636218, No. 11822403, No. 11804104, No. 11804105, No. 61875060 and No. U1801661), the Key R\&D Program of Guangdong province (Grant No. 2019B030330001), the Natural Science Foundation of Guangdong Province (Grant No.2015TQ01X715, No. 2014A030306012, No.2018A030313342 and No. 2018A0303130066), the Key  Project of Science  and  Technology of Guangzhou (Grant No. 201804020055).
\end{acknowledgments}

Y. M., Y. Z. and S. Z. contributed equally to this work.



\begin{thebibliography}{99}
\bibitem{Einstein1935}A. Einstein, B. Podolsky, and N. Rosen, Can quantummechanical description of physical reality be considered complete? Phys. Rev. \textbf{47}, 777 (1935).
\bibitem{Ou1992}Z. Y. Ou, S. F. Pereira, H. J. Kimble, and K. C. Peng, Realization of the Einstein-Podolsky-Rosen Paradox for continuous Variables, Phys. Rev. Lett. \textbf{68}, 3663 (1992).
\bibitem{Howell2004}J. C. Howell, R. S. Bennink, S. J. Bentley, and R. Boyd, Realization of the Einstein-Podolsky-Rosen Paradox Using Momentum- and Position-Entangled Photons from Spontaneous Parametric Down Conversion, Phys. Rev. Lett. \textbf{92}, 210403 (2004).
\bibitem{Lee2016}J.-C. Lee, K.-K. Park, T.-M. Zhao, and Y.-H. Kim, Einstein-Podolsky-Rosen Entanglement of Narrow-Band Photons from Cold Atoms, Phys. Rev. Lett. \textbf{117}, 250501 (2016).
\bibitem{Chen2019} L. Chen,T. Ma, X. Qiu, D. Zhang, W. Zhang and R. W. Boyd, Realization of the Einstein-Podolsky-Rosen Paradox Using Radial Position and Radial Momentum Variables, Phys. Rev. Lett. \textbf{123}, 060403(2019)
\bibitem{Bell1964}J. S. Bell, On the Einstein-Podolsky-Rosen paradox, Physics \textbf{1}, 195 (1964).
\bibitem{Clauser1969}  J. F. Clauser, M. A. Horne, A. Shimony, and R. A. Holt, Proposed Experiment to Test Local Hidden-Variable Theories, Phys. Rev. Lett. \textbf{23}, 880 (1969).
\bibitem{BoschiPRL1998} D. Boschi, S. Branca, F. De Martini, L. Hardy, S. Popescu, Experimental Realization of Teleporting an Unknown Pure Quantum State via Dual Classical and Einstein-Podolsky-Rosen Channels, Phys. Rev. Lett. \textbf{80}, 1121 (1998).
\bibitem{PanNP2010} X.-M. Jin, J.-G. Ren, B. Yang, Z.-H. Yi, F. Zhou, X.-F. Xu, S.-K. Wang, D. Yang, Y.-F. Hu, S. Jiang, T. Yang, H. Yin, K. Chen, C.-Z. Peng, J.-W. Pan, Experimental free-space quantum teleportation, Nat. Photon. \textbf{4}, 376 (2010).
\bibitem{BB84} C. H. Bennett and G. Brassard, Quantum cryptography: Public key distribution and coin tossing, In Proceedings of IEEE International Conference on Computers, Systems and Signal Processing, volume 175, page 8. New York, 1984.
\bibitem{QI} M. A. Nielsen and I. L. Chuang, Quantum Computation and Quantum Information. Cambridge University Press (2000).
\bibitem{ZhangOE2008}  Q. Zhang, H. Takesue, S. W. Nam, C. Langrock, X. Xie, B. Baek, M. M. Fejer, and Y. Yamamoto, Distribution of time-energy entanglement over 100 km fiber using superconducting single-photon detectors, Opt. Express \textbf{16}, 5776 (2008).
\bibitem{DuanPRL2000} L.-M. Duan, G. Giedke, J. I. Cirac, and P. Zoller, Inseparability Criterion for Continuous Variable Systems, Phys. Rev. Lett. \textbf{84}, 2722 (2000).
\bibitem{TombesiPRL2002} S. Mancini, V. Giovannetti, D. Vitali, and P. Tombesi, Entangling Macroscopic Oscillators Exploiting Radiation Pressure, Phys. Rev. Lett. \textbf{88}, 120401 (2002).
\bibitem{Jones2007}S. J. Jones, H. M. Wiseman, and A. C. Doherty, Entanglement, Einstein-Podolsky-Rosen correlations, Bell nonlocality, and steering, Phys. Rev. A \textbf{76}, 052116 (2007).
\bibitem{Wiseman2007}H. M. Wiseman, S. J. Jones, and A. C. Doherty, Steering, Entanglement, Nonlocality, and the Einstein-Podolsky-Rosen Paradox, Phys. Rev. Lett. \textbf{98}, 140402 (2007).
\bibitem{Cavalcanti2009}E. G. Cavalcanti, S. J. Jones, H. M. Wiseman, and M. D. Reid, Experimental criteria for steering and the Einstein-Podolsky-Rosen paradox, Phys. Rev.  A \textbf{80}, 032112 (2009).
\bibitem{SPDC} M. H. Rubin, D. N. Klyshko, Y. H. Shih, and A. V. Sergienko, Theory of two-photon entanglement in type-II optical parametric down-conversion, Phys. Rev. A \textbf{50}, 5122 (1994).
\bibitem{ReschPRL2018} J.-P. W. MacLean, J. M. Donohue, and K. J. Resch, Direct Characterization of Ultrafast Energy-Time Entangled Photon Pairs, Phys. Rev. Lett. \textbf{120}, 053601 (2018).
\bibitem{Wang2019} Y. Wang, J. Li, S. Zhang, K. Su, Y. Zhou, K. Liao, S. Du, H. Yan, and S.-L. Zhu, Efficient quantum memory for single-photon polarization qubits, Nat. Photon. \textbf{13}, 346 (2019).
\bibitem{DuJOSAB2008} S. Du, J. Wen, and M. H. Rubin, Narrowband biphoton generation near atomic resonance, J. Opt. Soc. Am. B \textbf{25}, C98 (2008).
\bibitem{DuPRL2008} S. Du, P. Kolchin, C. Belthangady, G.Y. Yin, and S. E. Harris, Subnatural Linewidth Biphotons with Controllable Temporal Length, Phys. Rev. Lett. 100, 183603 (2008).
\bibitem{Zhao2014} L. Zhao, X. Guo, C. Liu, Y. Sun, M. Loy, and S. Du, Photon pairs with coherence time exceeding one microsecond, Optica \textbf{1}, 84 (2014).
\bibitem{Zhang2012}S. Zhang, J. Chen, C. Liu, S. Zhou, M. Loy, G. K. L. Wong, and S. Du, A dark-line two-dimensional magneto-optical trap of 85Rb atoms with high optical depth, Rev. Sci. Inst. \textbf{83}, 073102 (2012).
\bibitem{Zhao2016}L. Zhao, Y. Su, and S. Du, Narrowband biphoton generation in the group delay regime, Phys. Rev. A \textbf{93}, 033815 (2016).
\bibitem{ClauserPRD1974} J. F. Clauser, Experimental distinction between the quantum and classical field-theoretic predictions for the photoelectric effect, Phys. Rev. D \textbf{9}, 853 (1974).
\bibitem{Grangier1986} P. Grangier, G. Roger, and A. Aspect, Experimental evidence for a photon anticorrelation effect on a beam splitter: a new light on single-photon interferences, Europhys. Lett. \textbf{1}, 173 (1986).
\bibitem{Zielnicki2018} K. Zielnicki, K. Garay-Palmett, D. Cruz-Delgado, H. Cruz-Ramirez, M. F. O'Boyle, B. Fang, V. O. Lorenz, A. B. U'Ren and P. G. Kwiat, Joint spectral characterization of photon-pair sources, J. Mod. Opt., \textbf{65}, 1141 (2018)
\end{thebibliography}
\end{document}